\documentclass[twocolumn,prd]{revtex4}
\usepackage{graphicx,float,alltt,psfig,curvesls,amsmath}

\begin{document}
\newcommand{\beq}{\begin{equation}} 
\newcommand{\eeq}{\end{equation}}
\newcommand{\bea}{\begin{eqnarray}} 
\newcommand{\eea}{\end{eqnarray}}
\newcommand{\om}{\Omega_{\rm M}}
\newcommand{\oll}{\Omega_{\rm \Lambda}}
\newcommand{\omol}{(\Omega_{\rm M}, \Omega_{\rm \Lambda})}
\newcommand{\bctr}{\begin{center}}
\newcommand{\ectr}{\end{center}}
\newcommand{\lsim}{\mbox{$\:\stackrel{<}{_{\sim}}\:$} }
\newcommand{\gsim}{\mbox{$\:\stackrel{>}{_{\sim}}\:$} }
\newcommand{\nl}{\newline}

\bctr 
{\large{\bf How far can the generalized second law be generalized?}}
\vspace{24pt}

P. C. W. Davies

{\em Australian Centre for Astrobiology, Macquarie University, Sydney, 2109, Australia pdavies@els.mq.edu.au}
\vspace{24pt}

Tamara M. Davis

{\em Dept. of Astrophysics, University of New South Wales,  Sydney, 2052, Australia \nl tamarad@phys.unsw.edu.au}
\ectr
\vspace{24pt}

{\small Jacob Bekenstein's identification of black hole event horizon area with entropy proved to be a landmark in theoretical physics. In this paper we trace the subsequent development of the resulting generalized second law of thermodynamics (GSL), especially its extension to incorporate cosmological event horizons. In spite of the fact that cosmological horizons do not generally have well-defined thermal properties, we find that the GSL is satisfied for a wide range of models. We explore in particular the case of an asymptotically de Sitter universe filled with a gas of small black holes as a means of casting light on the relative entropic `worth' of black hole versus cosmological horizon area. We present some numerical solutions of the generalized total entropy as a function of time for certain cosmological models, in all cases confirming the validity of the GSL.}

\vspace{44pt}
\section{Introduction}  

A key development in the history of physics came with Jacob Bekenstein's identification (Bekenstein, 
\cite{bekenstein73}) of entropy $S_{\rm bh}$ with black hole event horizon area $A_{\rm bh}$,
\beq S_{\rm bh}  \propto  k_{\rm B}\; A_{\rm bh}  \label{eq:SproptoA}\eeq
where $k_{\rm B}$ is Boltzmann's constant. That there was a close analogy between black hole event horizons and entropy was implied by Hawking's area theorem, which states that subject to certain reasonable physical conditions (most notably that energy cannot be negative) the total horizon area is a non-decreasing function of time (Hawking, 
\cite{hawking71}).  However, if a black hole were totally black it would have a zero temperature. Assuming the general relationship: entropy = energy/temperature, it would seem that the entropy of a black hole diverges. A similar conclusion follows from information theory. If the black hole forms from the implosion of a ball of matter, the information lost behind the event horizon is roughly $N$ bits, where $N$ is the number of particles in the ball. As classical physics imposes no lower bound on the mass of a particle, there is no upper bound on $N$. Then identifying entropy with information loss confirms that $S_{\rm bh}$ diverges.

It was Bekenstein's suggestion, with the encouragement of John Wheeler, that quantum mechanics removes the divergence by placing a lower bound on the mass of the particles that go to make up the black hole (Bekenstein, 
\cite{bekenstein73}). In order to confine a particle to the Schwarzschild radius, its Compton wavelength must be less than the size of the hole, from which Bekenstein concluded that the constant of proportionality in Eq.~\ref{eq:SproptoA} includes the factor $1/\hbar$. These essential ideas were later confirmed following the application of quantum field theory to the formation of black holes by Hawking 
\cite{hawking75}, 
 who discovered that a Schwarzschild black hole of mass $M_{\rm bh}$ and surface gravity  $\kappa_{\rm bh}$ radiates with the temperature
\bea T_{\rm bh} &=& \hbar/(2\pi k_{\rm B}c)\kappa_{\rm bh}\\ 
&=&\frac{\hbar c^3}{8\pi k_{\rm B} G M_{\rm bh}}.\eea
This fixes the constant of proportionality in Eq.~\ref{eq:SproptoA},
\bea S_{\rm bh} &=& \frac{k_{\rm B}c^3}{4G\hbar} A_{\rm bh}  \\
                &=& \frac{1}{4}A_{\rm bh} \label{eq:A4}\eea
where Eq.~\ref{eq:A4} uses units with $\hbar = c = G = k_{\rm B} = 1$. We shall adopt these units henceforth. 

Following these developments, the second law of thermodynamics could then be generalized to include cases where black holes exchange heat and energy with their environment (Hawking, \cite{hawking75,hawking76}),
\beq \dot{S}_{\rm bh} + \dot{S}_{\rm m} \ge 0\eeq
where $S_{\rm m}$ is the entropy of the matter and an overdot represents differentiation with respect to proper time. The fact that black holes radiate implies that they can lose energy and shrink, in violation of Hawking's area theorem. The strictures of the theorem are evaded because the quantum state permits a flux of negative energy into the hole (Davies, Fulling, and Unruh, 
 \cite{davies76}). 
However, the thermal radiation emitted by the black hole always raises the entropy of the environment by at least as much as the loss of horizon  entropy caused by the shrinkage. The generalized second law (GSL) was extended to black holes with rotation and electric charge in a straightforward manner (Hawking, 
\cite{hawking75}).

\vspace{24pt}
\section{Extending the generalised second law of thermodynamics to cosmological horizons}

Shortly after the establishment of black hole horizon entropy, it was proposed that the GSL be extended to include the event horizon area of de Sitter space (Gibbons \& Hawking, 
\cite{gibbons-hawking77}).  This cosmological model has attracted interest in the last twenty years because, if the inflationary scenario is correct, the very early universe experienced a period of exponential expansion that approximated a portion of de Sitter space.  Moreover, recent cosmological observations suggest that the cosmological constant $\Lambda$ is non-zero and positive (see Lineweaver 
\cite{lineweaver01} and Sievers et al. 
\cite{sievers02} for recent reviews) --- all ever-expanding Friedmann-Robertson-Walker (FRW) universes of this class tend towards de Sitter space at late times. Hence universes with event horizons have recently experienced a resurgence of interest.  
 
De Sitter space has a time-independent event horizon at the proper distance $r_{\rm c} = 1/H = \sqrt{3/\Lambda}$ (subscript $c$ stands for {\em cosmological} event horizon).  Given Hubble's law $v=Hr$, a simple rearrangement shows that the event horizon is at the distance at which, using the above definition of $v$, comoving test particles recede at the speed of light.  In a de Sitter universe particles receding faster than light are beyond our view (but not in general for FRW universes (Kiang, 
\cite{kiang97})). 

\begin{figure}\bctr
\includegraphics[width=86mm]{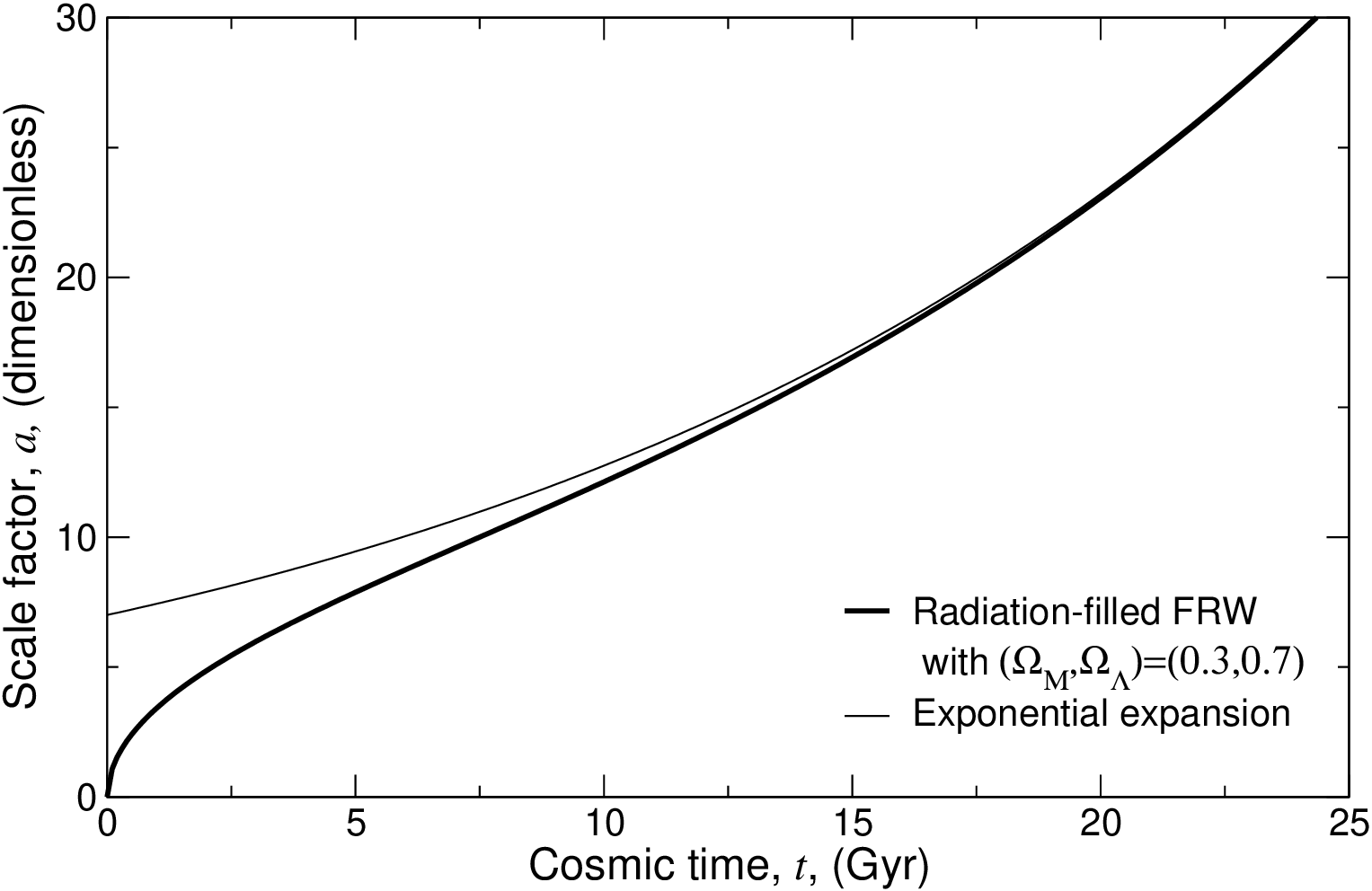} 
\caption{\small{Scale factor is plotted against time for a flat, radiation filled, FRW universe (Eq.~\ref{eq:at}).  Hubble's constant is taken to be $H_0=70 {\rm kms}^{-1}{\rm Mpc}^{-1}$, and the cosmological parameters $\Omega_M=8\pi G\rho_{\rm 0}/3H_0^2$ and $\Omega_{\Lambda}=\Lambda/3H_0^2$ are given their observationally favoured values 0.3 and 0.7 respectively.  The expansion initially decelerates while the gravity of the radiation dominates, then begins to accelerate as the cosmological constant takes over.  In this universe, as in all eternally expanding FRW universes with non-zero $\Lambda$, the expansion tends toward exponential at late times, $a(t) \propto e^{Ht}$.   }}
\label{fig:rad-FRW}
\ectr\end{figure}

The thermal properties of de Sitter space may be deduced from the fact that the vacuum Green function for conformally-invariant fields propagating in a de Sitter background spacetime are periodic in imaginary time with a periodicity,  $2\pi/\kappa_{\rm c}$, corresponding to a thermal state with temperature
\bea T_{\rm deS} &=& \kappa_{\rm c}/2\pi, \label{eq:TdeSitt}\\
&=& \frac{1}{2\pi}\left(\frac{\Lambda}{3}\right)^{1/2} \eea
where $\kappa_{\rm c}=(\Lambda/3)^{1/2}$ is the surface gravity of the cosmological event horizon. The same result may be established by considering the response of a particle detector travelling along a geodesic in a de Sitter-invariant vacuum state; the detector behaves as if immersed in a bath of thermal radiation with temperature given by Eq.~\ref{eq:TdeSitt}.  
Gibbons and Hawking 
\cite{gibbons-hawking77} therefore made the reasonable suggestion that the GSL be extended to de Sitter space. (For more general derivations of horizon entropy see the recent pioneering work of Padmanabhan 
\cite{padmanabhan02a,padmanabhan02b,padmanabhan02c}.)
There were, however, some considerations that made the identification of de Sitter event horizon area with entropy less compelling than in the black hole case.  First, de Sitter horizons do not radiate in the manner of black holes. In spite of the thermal response of a particle detector, the expectation value of the stress-energy momentum tensor of the de Sitter-invariant massless scalar field vacuum state is not that of a bath of thermal radiation.  It is given instead by (see, for example, Birrell and Davies, 
\cite{birrell82})
\beq T_{\mu}\,^{\nu} = \frac{\Lambda^4}{8640\pi^2}\; \delta_{\mu}\,^{\nu}.\eeq
This  corresponds to the stress-energy-momentum tensor associated with a cosmological constant, and so merely renormalizes $\Lambda$.  Second, in the black hole case the observer lies outside the horizon, whereas in the cosmological case the horizon not only envelops the observer, its location is observer-dependent.  There is therefore no apparent source of the radiation; it appears homogeneous.  Third, there is no parameter in de Sitter space that corresponds to the mass of the black hole, making it problematic to balance the books in a trade-off of energy and entropy between the de Sitter horizon and any matter present.  Finally, a key heuristic in Bekenstein's derivation of Eq.~\ref{eq:SproptoA} was the association of information with entropy.  The black hole irreversibly swallows the information of the body that implodes to form it, and the entropy of the horizon may be thought of as a measure of this lost information.  In the case of de Sitter space, the region beyond the horizon is infinite, and can accommodate an infinite quantity of information.  

In spite of these shortcomings, we shall show that the GSL can be consistently applied to de Sitter horizons that exchange heat and energy with both other horizons and with matter.  

To investigate the entropy and energy budgets when the de Sitter horizon exchanges heat with other systems, we first consider the case of a FRW universe with a positive $\Lambda$ containing uniform thermal radiation with temperature $T > T_{\rm deS}=H/2\pi$ (which corresponds to a typical wavelength of the radiation being less than the horizon radius). Consider a FRW universe model with scale factor $a$, and Robertson-Walker metric
\beq   ds^2 = dt^2 -a^2(t) \left[d\chi^2+ S_k(\chi)^2(d\theta^2 + \sin^2\theta d\phi^2)\right]  \label{eq:RW}\eeq 
where $S_k(\chi) = \sinh \chi, \chi, \sin \chi$ for $k=-1,0,1$ respectively. The Friedmann equations may be written
\bea \dot{\rho} &=& -3H(\rho + p)\label{eq:Fried1}\\
3H^2&=&8\pi\rho + \Lambda - 3k/a^2\label{eq:Fried2}\eea 
where $\rho$ and $p$ are the density and pressure of the cosmological fluid respectively, and we take $p = \rho/3$ for radiation.  In the spatially flat case, $k = 0$, Eqs.~\ref{eq:Fried1} and \ref{eq:Fried2} may readily be solved to give the scale factor
\beq a(t)  =  \left(\frac{8\pi\rho_{\rm 0}}{\Lambda}\right)^{1/4}\sinh^{1/2}\left(2t\sqrt{\frac{\Lambda}{3}}\right) \label{eq:at}\eeq 
where $\rho_{\rm 0}$ is the present day radiation density and  $\rho=\rho_0 a^{-4}$.   Eq.~\ref{eq:at} is plotted in Fig.~\ref{fig:rad-FRW}, from which it can be seen that the universe starts out with a standard hot big bang, and approaches de Sitter space as $t\rightarrow \infty$.

Consider a small departure from de Sitter space, such that the distance to the cosmological horizon, $r_{\rm c}$, is approximately $1/H$, corresponding to an entropy of $S_{\rm c}\approx\pi/H^2$.  This will be a good assumption at late times.  The rate of change of cosmological horizon entropy is then approximately
\beq \dot{S}_{\rm c}\approx -2\pi\dot{H}/H^3 = 2\pi (16\pi\rho)/3H^3 \label{eq:Hdot}\eeq
using Eq.~\ref{eq:Fried2}. This rise in horizon entropy will be traded against a loss of radiation entropy as the radiation streams away across the de Sitter horizon.  The radiation energy density can be expressed as $\rho=\sigma T^4$ where the radiation constant $\sigma =\pi^2/15$ in these units. The entropy density is $s=(4/3)\rho T^{-1}$, so the total radiation entropy within a horizon volume is given by
\beq S_{\rm r} = \frac{16\pi \sigma^{1/4}}{9}\; \rho^{3/4}H^{-3}. \eeq 
Using Eq. (11), the rate of loss of radiation entropy is 
\bea \dot{S}_{\rm r} &=& \frac{16\pi \sigma^{1/4}}{9}(3/4 \rho^{-1/4} \dot{\rho} H^{-3} - 3\rho^{3/4}\dot{H}H^{-4})\label{eq:ent}\\
                     &\approx&  \frac{-16\pi\sigma^{1/4}}{3} \; \rho^{3/4}H^{-2} \label{eq:Srad}\eea
if $\dot{H}$ is small.  The condition for the second law to be satisfied is 
\beq \dot{S}_{\rm c} + \dot{S}_{\rm r} \ge 0. \label{eq:S}\eeq
Comparing Eq.~\ref{eq:Srad} with Eq.~\ref{eq:Hdot} one sees that the gain in horizon entropy exceeds the loss of radiation entropy if $\rho^{1/4} > H/2\pi$. This is just the condition that the radiation temperature be greater than the horizon temperature, as assumed.

\begin{figure}\bctr
\includegraphics[width=86mm]{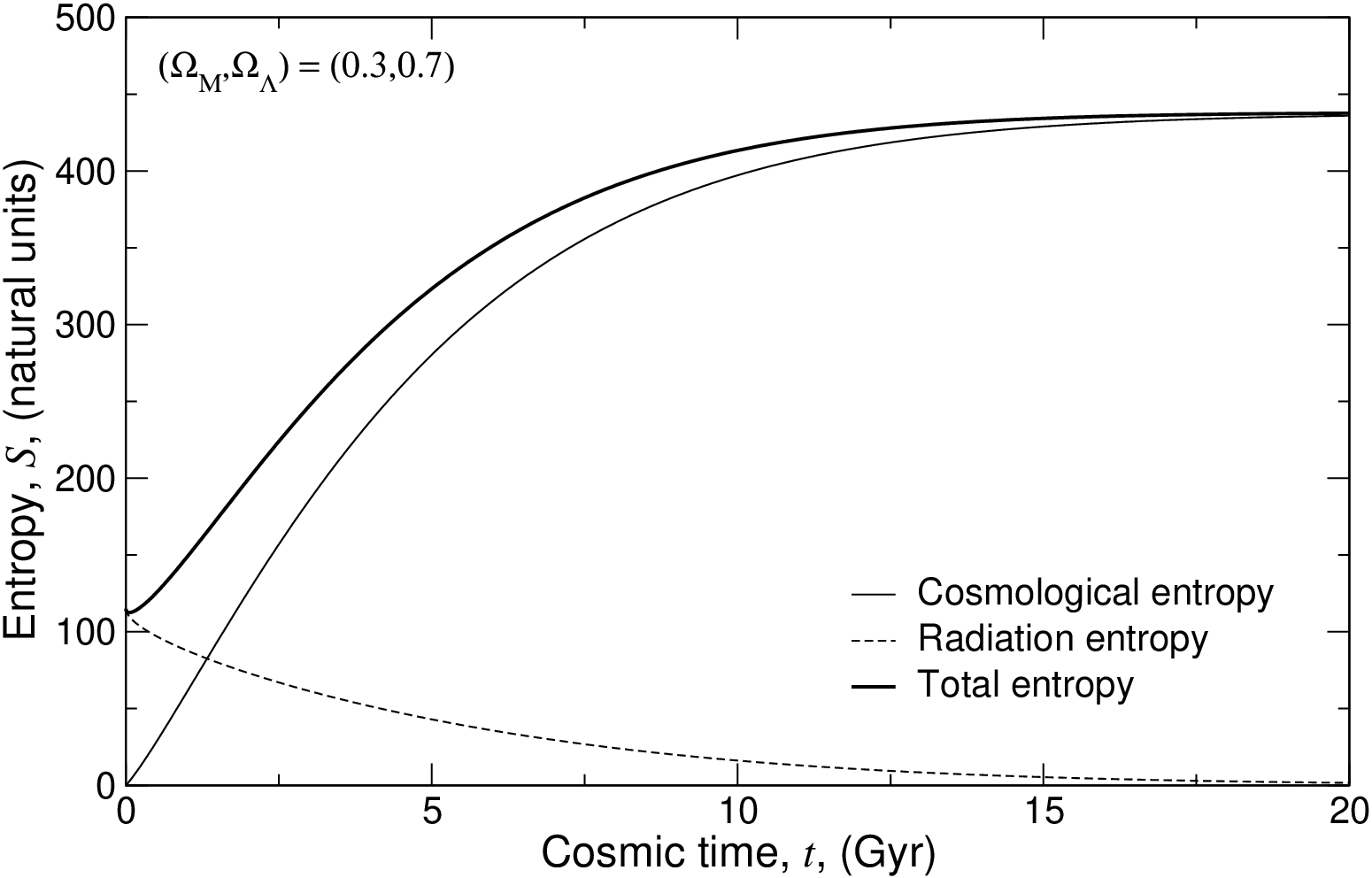} 
\caption{\small{The entropy of the cosmological horizon, the entropy of the radiation within the horizon and the combined total entropy are plotted against time for a radiation dominated FRW model with $\omol=(0.3,0.7)$.  At all times the loss of radiation entropy across the cosmological horizon is less than the increase in cosmological horizon entropy, so the GSL is satisfied.}}
\label{fig:rc}
\ectr\end{figure}

In the foregoing we have assumed a small departure from de Sitter space.  Unfortunately when we relax this condition the distance to the event horizon cannot be solved exactly, but we show a numerical solution in Fig.~\ref{fig:rc}.  From this it may be seen that the horizon entropy always rises faster than the loss of radiation entropy, confirming the validity of the GSL even at early times.  In arriving at these results we define the (time-dependent) horizon radius as
\beq r_{\rm c} = a(t)\int_t^\infty \frac{dt^\prime}{a(t^\prime)}. \label{eq:rc}\eeq  

It is possible to prove an analogue of the Hawking area theorem for cosmological horizons (Davies, 
\cite{davies88a}). Recall that it was this theorem that originally  motivated the association of entropy with black hole horizon area. Consider the class of homogeneous isotropic models filled with a fluid of pressure $p$ and energy density $\rho$ and metric given by Eq.~\ref{eq:RW}.
It is convenient to work with the conformal time parameter defined by 
\beq \eta = -\int_t^\infty \frac{dt^\prime}{a(t^\prime)}.\label{eq:eta}\eeq
If $|\eta|<\infty$ for a particular model universe, then that universe possesses an event horizon.  

Comparing Eq.~\ref{eq:rc} and Eq.~\ref{eq:eta} we see that the distance to the cosmological event horizon is given by $r_{\rm c}= -a\eta$.  When the area of the event horizon is calculated in the $k = \pm 1$ cases we must take into account curvature terms. We prefer to work with the corrected distance
\beq r_{\rm c}  = -a S_k(\eta).\label{eq:Rh}\eeq
With this definition, the area of the event horizon always increases if 
\beq r^\prime_{\rm c}\ge0 \label{eq:Rprime}\eeq
 where a prime indicates differentiation with respect to $\eta$.  Performing this differentiation on Eq.~\ref{eq:Rh} gives the condition
\beq  \dot{a} \ge -1/T_k(\eta) \label{eq:Tk}\eeq
where $T_k(\chi) = \tanh \chi, \chi, \tan \chi$ for $k=-1,0,1$ respectively. It is interesting to give a physical interpretation of Eq.~\ref{eq:Tk}. For the flat ($k=0$) case it implies that the event horizon area does not decrease as long as $\dot{a}\ge-1/\eta$, or $r_{\rm c} \ge 1/H$.  That is, the event horizon area always increases as long as it is farther away than the Hubble sphere (the distance at which the recession velocity, using the definition $v=Hr$, is equal to the speed of light).  Note that the inequality allows the event horizon to be {\rm more} distant than the Hubble sphere: in most FRW universes we can observe objects which are receding faster than the speed of light.  The condition is slightly varied for the curved cases, because in these cases it is sometimes possible for objects receding less than the speed of light to be beyond the event horizon. 

We now assume that the cosmological fluid is subject to the dominant energy condition $\rho + p \ge 0$. This is the same condition as used in the black hole area theorem --- the one that is violated by quantum effects, allowing black holes to evaporate. It then follows from Eq.~\ref{eq:Fried1} that $\dot{\rho}\le 0$, from which inequality Eq.~\ref{eq:Rprime} readily follows (Davies, 
\cite{davies88a}).

In the cosmological case, there is no direct analogue of the evaporation of a black hole. However, one may relax the dominant energy condition by considering cosmological fluids with pressure $p=(\gamma-1)\rho$ and bulk viscosity $\alpha \rho >0$. It is then possible (Davies, 
\cite{davies87}) to have a combined effective pressure of the fluid plus the cosmological contant in excess of minus the combined energy density.
Equation~\ref{eq:Hdot} is then modified to 
\beq \dot{S}_{\rm c} = 8\pi^2 (\gamma -3H\alpha)\rho/H^3 \label{eq:8pi}\eeq
from which it follows that the horizon entropy will decrease if $\gamma < 3H\alpha$.  This does not, however, necessarily signal a violation of the GSL.
The presence of viscosity implies the generation of heat entropy.  A simple calculation (Davies, 
\cite{davies87}) shows that the fluid generates heat entropy at a rate 
\beq \dot{S}_{\rm m} = \frac{9H^2\alpha \rho a^3}{T}\label{eq:Sdot2}\eeq
where $T$ is the temperature of the fluid and we identify $a^3$ at an instant of cosmic time, $t$, with a horizon volume.  Comparing Eq.~\ref{eq:8pi} and Eq.~\ref{eq:Sdot2} for the case $\gamma=0$ (corresponding to minimal departure from the condition $\rho + p =0$) shows that the total entropy is unchanged for the equilibrium case
\beq T= \frac{H}{2\pi} = T_{\rm deS} \eeq
thus respecting the GSL.  The case $T>T_{\rm deS}$ requires a more complicated treatment (Davies, 
\cite{davies87}) but also conforms with the GSL.  Several exact solutions of cosmological models with viscosity have been discussed in detail by Barrow 
\cite{barrow87} 
and Davies 
\cite{davies88b}.

\section{Black hole-de Sitter spacetimes}
Special interest attaches to cosmological models that contain both black holes and cosmological horizons. As these spacetimes evolve, there may be an exchange of horizon area between the two. The question then arises of whether cosmological horizon area has the same `worth' as black hole horizon area. If the loss of black hole horizon area was not compensated by at least as great gain in cosmological horizon area, for example, then the GSL would be threatened.

A general class of black-hole--de Sitter solutions including mass $M$, charge $Q$, and angular momentum $J$ was provided by Carter 
\cite{carter73}.  The horizons for these solutions fall at the roots of the quartic equation 
\beq r^4 + \left(\frac{1}{\Lambda}-J^2\right)r^2 + \frac{2Mr}{\Lambda}-\left(\frac{J^2+Q^2}{\Lambda}\right)=0. \label{eq:r4}\eeq
If the four roots are given by $r_{++}>r_{+}>r_{-}>r_{--}$ then $r_{++}$ is identified with the cosmological event horizon, and $r_{+}$ with the outer horizon of the black hole.  For an observer in the region outside the hole, but within the cosmological event horizon, the total horizon area is 
\beq A_{\rm tot}=4\pi (r_{++}^2 + r_{+}^2 + 2J^2)\label{eq:Ainit}\eeq
(Gibbons and Hawking, 
\cite{gibbons-hawking77}). 
This spacetime is thermodynamically unstable: the black hole temperature is always greater than the cosmological horizon temperature. The black hole slowly evaporates, and the resulting radiation eventually passes over the cosmological horizon. At late times the solution settles down to de Sitter, with horizon area  
\beq A_{\rm deS} = 12\pi \Lambda^{-1}. \label{eq:AdeS}\eeq
It follows from the algebraic conditions on the four roots of the quartic (Eq.~\ref{eq:r4}) that
\beq   \Sigma_{i=1}^4\; r_i\,^2 = 2 \left(\frac{1}{\Lambda^2} - a^2\right) \eeq
from which it readily follows that Eq.~\ref{eq:AdeS} is greater than Eq.~\ref{eq:Ainit}, in conformity with the GSL  (Davies, 
\cite{davies84}). For black holes with charge $Q$ and angular momentum $J$ the proof is limited to the parameter range $\Lambda^{-1}>3M^2>3J^2$.  This is roughly the range for which the cosmological horizon lies outside the black-hole horizon and the central mass is not a naked singularity.  

Davies, Ford and Page 
\cite{davies-etal86} extended the foregoing `before-and-after' analysis by considering the continuous exchange of small quantities of heat between the black hole and the cosmological horizon. They introduced a further generalization by considering a massive shell around the black hole. This has the effect of depressing the temperature of the hole. A sufficiently massive shell lying sufficiently close to the black hole horizon can produce a backflow of heat from the cosmological horizon into the black hole. It turns out that even in these circumstances the GSL is satisfied.

As a final refinement of the investigation of the trade-off between black hole and cosmological horizons, we consider a FRW model universe with positive cosmological constant and a fluid consisting of uniform dust composed entirely of identical slow-moving small Schwarzschild black holes of mass $M_{\rm bh}$ and number density $n_{\rm bh}$. We ignore the Hawking effect (it could be suppressed if necessary by considering maximally charged black holes) and the small correction to the black hole radius implied by a non-zero $\Lambda$.   
In this case the Friedmann equations, Eq.~\ref{eq:Fried1} and Eq.~\ref{eq:Fried2}, can be written 
\bea \dot{\rho}_{\rm bh}&=& -3H\rho_{\rm bh}\label{eq:Fried1b} \\
3H^2 &=& 8\pi \rho_{\rm bh} +\Lambda -3k/a^2\label{eq:Fried2b}\eea
where $\rho_{\rm bh}=M_{\rm bh} \; n_{\rm bh}$.	

Restricting for the moment to small perturbations about de Sitter space, we may proceed in the same fashion as the calculation in Eqs.~\ref{eq:Hdot}-\ref{eq:S} 
 above.  Eq.~\ref{eq:Hdot} gives the rate of change of the cosmological horizon entropy, $\dot{S}_{\rm c} = \dot{A}_{\rm c}/4$. The total black hole area within the cosmological horizon is, 
\bea A_{\rm bh}^{\rm total} &=& \; A_{\rm bh} \; n_{\rm bh} V_{\rm c}\\ 
&=& \frac{64\pi^2 M_{\rm bh}\rho_{\rm bh}}{3\,H^3}.\eea
The rate of change of black hole area is therefore,
\beq \dot{A}_{\rm bh}^{\rm total}= -\frac{64\pi^2 M_{\rm bh} \rho_{\rm bh}}{H^2}\label{eq:Atotbh} \eeq
where we have used Eqs.~\ref{eq:Fried1b} and~\ref{eq:Fried2b}, and once again ignored the $\dot{H}$ term.  For the GSL to hold we need,
\beq \dot{A}_{\rm bh}^{\rm total} + \dot{A}_{\rm c} \ge 0. \label{eq:Abhdot}\eeq
Using Eqs.~\ref{eq:Hdot} and~\ref{eq:Atotbh} this inequality becomes 
\bea 2M_{\rm bh} &\lsim & 1/H\\
                  r_{\rm bh} &\lsim & r_{\rm c}.\eea 
Thus the GSL holds as long as the black holes are smaller than the cosmological event horizon.

\begin{figure}[t]\bctr
\includegraphics[width=86mm]{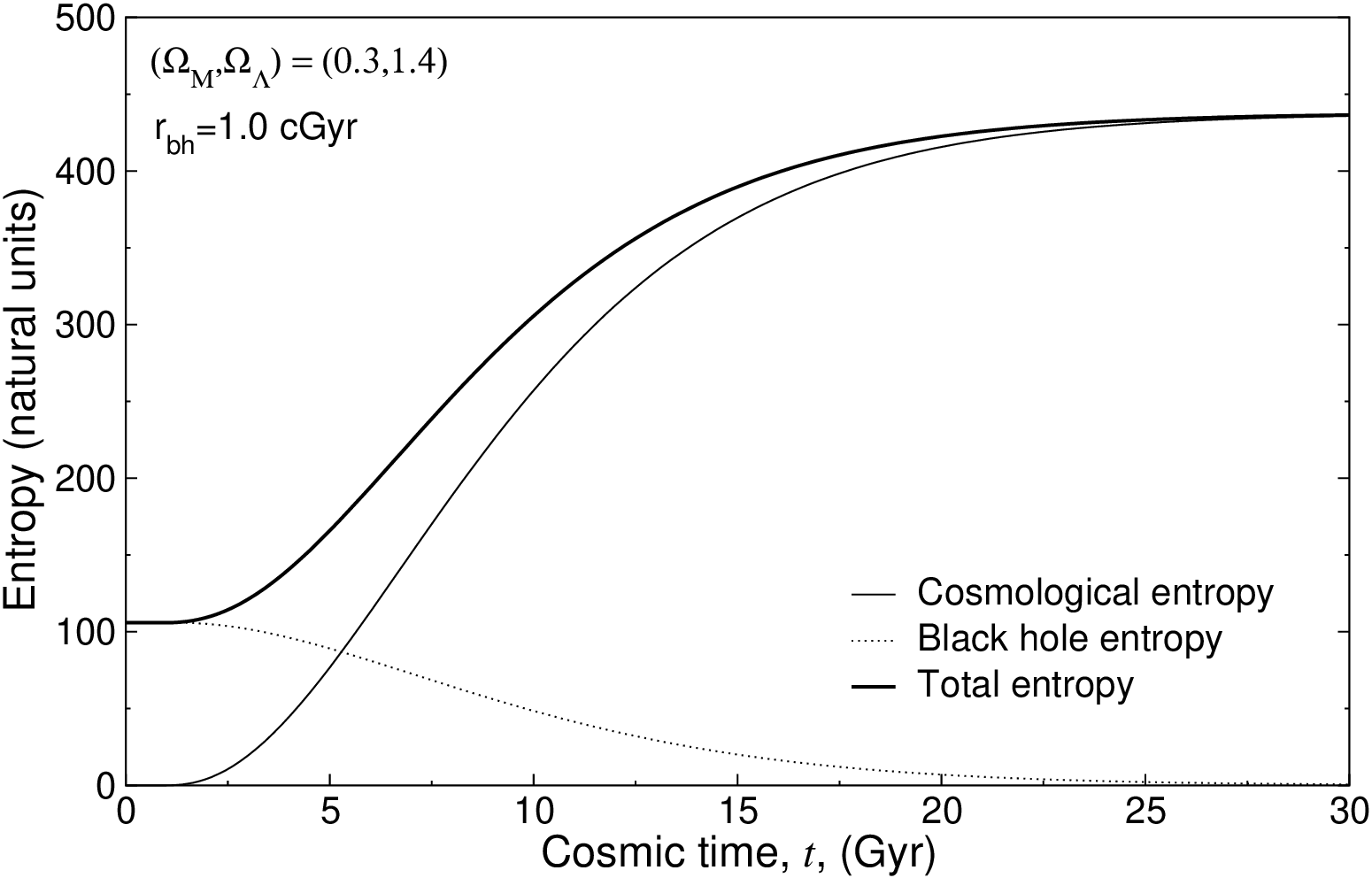} 
\caption{\small{Entropy variation as black holes cross the cosmological event horizon.  The model shown is a FRW universe with $\omol=(0.3,1.4)$ in which the matter density consists entirely of a comoving gas of black holes.  This is an example of a closed universe, and the cosmological event horizon appears out of the antipode at a finite time --- prior to this the black hole entropy is constant and the cosmological horizon entropy is zero.}} 
\label{fig:rbh10}
\ectr\end{figure}

We have been unable to solve inequality~\ref{eq:Abhdot} exactly for all FRW universes, but we give a sample numerical solution for large departures from de Sitter space in Fig.~\ref{fig:rbh10}.  Note that the total entropy is a monotonically increasing function of cosmic time $t$ at all times.  We shall report elsewhere (Davis et al. 
\cite{davis03event}) on a broader range of numerical solutions of black hole cosmological models.

\section{Relation to other work}
In this review we have concentrated on several explicit examples, including some numerical solutions, of extensions of Bekenstein's generalized second law of thermodynamics to cosmological horizons. 
This work complements some recent theorems that prove more general but less explicit results that have a bearing on the GSL. For example, for black hole-de Sitter spacetimes, Shiromizu et al. 
\cite{shiromizu93} 
show that the black hole event horizon area is non-decreasing in asymptotically de Sitter space times,  
while Hayward et al. 
\cite{hayward94} 
show that the black hole event horizon area is bounded by $4\pi/\Lambda$.

The cosmological horizon area in models with a positive cosmological constant has been considered by Boucher et al. 
\cite{boucher84} 
who show that the horizon area is bounded by $12\pi/\Lambda$ on a regular time-symmetric hypersurface, while
Shiromizu et al. 
\cite{shiromizu93} 
show that the horizon area is bounded by $12\pi/\Lambda$ on a maximal hypersurface.
Neither of these proves a bound in a non-stationary asymptotically de Sitter universe such as a FRW universe.

Maeda et al 
\cite{maeda97}
extend the work of Davies 
\cite{davies88a} by showing
that the cosmological event horizon area doesn't decrease in any asymptotically de Sitter spacetimes.  They also show that the de Sitter horizon is the upper limit of horizon size for any cosmological model with nonzero $\lambda$. Our results both illustrate these theorems and demonstrate that, for certain specific models, the GSL is satisfied not just asymptotially, but at all times.

\vspace{32pt}
{\bf Postscript.}

The reference to Padmanabhan 
\cite{padmanabhan02a,padmanabhan02b,padmanabhan02c}
 was inadvertantly omitted from the published version.

{\bf Acknowledgements.}
We should like to thank Charles Lineweaver for helpful discussions.   TMD is supported by an Australian Postgraduate Award.

\appendix
\renewcommand\baselinestretch{1.2}
\normalsize

\end{document}